\documentclass{ws-procs9x6-cpt25}
\begin{document}

\newcommand{\refeq}[1]{(\ref{#1})}
\def\etal {{\it et al.}}

\title{Cosmic Backgrounds in the \\ Gravitational Standard-Model Extension}

\author{C.M.\ Reyes,$^1$ C.\ Riquelme,$^{1,2}$ M.\ Schreck,$^{3}$, and A.\ Soto$^4$}

\address{$^1$Centro de Ciencias Exactas, Departamento de Ciencias B\'{a}sicas, Facultad de Ciencias, Universidad del B\'{i}o-B\'{i}o, Chill\'{a}n, Casilla 447, Chile }

\address{$^2$Departamento de F\' {\i}sica, Universidad de Concepci\'on \\ Casilla 160-C, Concepci\'on, Chile }

\address{$^3$Departamento de F\'{i}sica, Universidade Federal do Maranh\~{a}o, Campus Universit\'{a}rio do Bacanga, S\~{a}o Lu\'{i}s (MA), 65085-580, Brazil }

\address{$^4$School of Mathematics, Statistics and Physics, Newcastle University \\ Newcastle upon Tyne, NE1 7RU, UK}

\begin{abstract}
We consider background fields within the gravitational sector of the 
Standard-Model Extension (SME) in a cosmological setting. Our analysis is divided into two parts. 
The first part addresses the consistency of nondynamical backgrounds in scenarios 
where diffeomorphism invariance is explicitly broken. Focusing on gravitational 
systems that admit Killing vector fields and possess \textit{a priori} symmetries, we 
demonstrate that potential discrepancies between Riemannian geometry and 
dynamical equations can be avoided. The second part presents a direct application 
of various techniques developed by decomposing the modified Einstein equations 
along normal and tangential directions of the $(3+1)$ decomposition.
We show that nondynamical backgrounds can lead to accelerated cosmological expansion without requiring a
cosmological constant, thereby opening new avenues for interpreting dark energy.
\end{abstract}

\bodymatter

\section{Introduction}
The cosmological Standard Model $\Lambda$CDM provides 
a remarkably precise description of the
universe's evolution. It accurately accounts for the cosmic microwave background (CMB), 
primordial nucleosynthesis, and the
formation of large-scale structures. Despite its success, the model faces several limitations,
including the unknown nature of dark matter, the mechanism behind late-time accelerated
expansion and the lack of a fundamental description of physics in the pre-inflationary epoch.

Background fields in gravity have been proposed to unveil potential
effects of diffeomorphism violation;\cite{Kostelecky:2003fs} see also Ref.~\refcite{deRham:2010kj}.
Recently, both dynamical and nondynamical backgrounds have been considered
as viable candidates for dark energy.\cite{Reyes:2024hqi,Nilsson:2023exc,Reyes:2022dil}
These developments are particularly compelling in the light of recent 
DESI results,\cite{DESI:2025zgx,DESI:2025fii} which, when combined with CMB and supernovae data,
suggest that accelerated expansion may be better explained by a
field evolving in time rather than a cosmological constant.
\section{Dynamical and nondynamical backgrounds}
On the one hand, the gravitational sector of the minimal SME with explicit diffeomorphism violation 
can be written in the form
\begin{equation}\label{Actionprin}
S_{\text{eSME}}=\int_{\mathcal{M}}   \mathrm{d}^4x\,\frac{\sqrt{-g}}{2\kappa}  
 \left(  R+2\Lambda+\mathcal L_{ust} \right)+S_{ \partial \mathcal M} +S_{ \text{matter}}   \,.
\end{equation}
Here $\kappa = 8\pi G$, with $G$ denoting Newton’s gravitational constant, $g$ is the
determinant of the metric $g_{\mu \nu}$, and $\Lambda$
the cosmological constant. The 
 Lagrangian $\mathcal{L}_{ust}$ is usually
 decomposed into three parts,
\begin{equation}
\mathcal L_{ust}=-u {R}+s^{\mu \nu}{R}_{\mu \nu}+t^{\mu\nu\rho\sigma}{R}_{\mu\nu\rho\sigma}\,,
\end{equation}
where $u$, $s^{\mu \nu}$, and $t^{\mu \nu \rho \sigma}$ are nondynamical background fields that couple to 
the 
 Ricci scalar $R$, the Ricci tensor $R_{\mu \nu}$ and the Riemann tensor $R_{\mu \nu \rho \sigma}$, respectively.
By construction, these backgrounds do not exhibit fluctuations, which is 
expressed mathematically by $\delta X = 0$ with $X \in\{ u, s^{\mu\nu}, t^{\mu\nu\varrho\sigma}\}$.
The piece $S_{\text{matter}}$ accounts for a matter contribution, which will be specified later.

The action term $S_{\partial \mathcal{M}}=\sum_{i=u,s,t} S_{   \partial \mathcal{M}  }^{({i})}$ contains
 boundary terms to ensure a well-posed variational principle.\cite{Reyes:2021cpx,Reyes:2024hqi}
 They read
 \begin{subequations}
 \begin{align}
S_{   \partial \mathcal{M}  }    ^{({u})}   &=-\oint_{\partial\mathcal{M}} \mathrm{d}^3y\,
  \varepsilon \frac{\sqrt{q}}{\kappa}\, uK   \,,
\\[1ex]
S_{   \partial \mathcal{M}  }    ^{({s})}   &=  \oint_{\partial\mathcal{M}} \mathrm{d}^3y\, 
 \varepsilon \frac{\sqrt{q}}{2\kappa}\,\left( s^{ab}K_{ab}-s^{\mathbf{nn}} K\right)  \,,
\\[1ex]
S_{   \partial \mathcal{M}  }    ^{({t})}   &= \oint_{\partial\mathcal{M}}  
 \mathrm{d}^3y \, \varepsilon \frac{\sqrt{q}}{\kappa} \,   2 t^{\mathbf{n}a\mathbf{n}b} K_{ab}   \,.
\end{align}
\end{subequations}
Here, $q_{ab}$ is the induced metric on the boundary hypersurface $\partial \mathcal{M} $, $q$ its determinant, and
$\varepsilon=n_{\mu} n^{\mu}$ where $n_{\mu}$ is the boundary normal evaluated at each point of $\partial\mathcal{M}$.
We have introduced
 the extrinsic curvature $K^{ab}$, its trace $K$, and 
the notation $X^{\bf {n}}:=n_{\mu }X^{\mu}$.
 
Varying the action of Eq.~\refeq{Actionprin} leads to the modified Einstein field equations\cite{Bailey:2006fd}
\begin{equation}\label{EM}
    G_{\mu\nu}+\Lambda g_{\mu\nu}=(T^{Rstu})_{\mu\nu}+\kappa(T_m)_{\mu\nu}\,,
\end{equation}
where $G_{\mu \nu}=R_{\mu \nu}-(R/2)g_{\mu\nu}$ is the Einstein tensor and 
$(T_m)_{\mu\nu}$ denotes the energy-momentum tensor of the matter content.
The contribution from the background fields is represented by the second-rank tensor
\begin{align}
    2(T^{Rstu})^{\mu\nu}&=-\nabla^\mu \nabla^\nu u-\nabla^\nu \nabla^\mu u+2g^{\mu\nu}\nabla^2 u+2u G^{\mu\nu}
    +s^{\alpha\beta}R_{\alpha\beta}g^{\mu\nu}
     \notag \\    
     &\phantom{{}={}}+\nabla_\alpha \nabla^\mu s^{\alpha\nu}+\nabla_\alpha \nabla^\nu s^{\alpha\mu} -\nabla^2 s^{\mu\nu}
    -g^{\mu\nu}\nabla_\alpha \nabla_\beta s^{\alpha\beta} \notag \\
    &\phantom{{}={}}+t^{\alpha\beta\gamma\mu}R_{\alpha\beta\gamma}^{\phantom{\alpha\beta\gamma}\nu}
    +t^{\alpha\beta\gamma\nu}R_{\alpha\beta\gamma}^{\phantom{\alpha\beta\gamma}\mu}
    +t^{\alpha\beta\gamma\delta}R_{\alpha\beta\gamma\delta}g^{\mu\nu}   \notag  \\
    &\phantom{{}={}}-2\nabla_\alpha\nabla_\beta t^{\mu\alpha\nu\beta}  -2\nabla_\alpha\nabla_\beta t^{\nu\alpha\mu\beta}  \,.
\end{align}
On the other hand, spontaneous spacetime symmetry breaking is
 introduced through a dynamical field acquiring a spacetime-dependent vacuum 
 expectation value.\cite{Bluhm:2004ep,Reyes:2024hqi,Santos:2024iyc} A widely studied setting is the bumblebee model:\cite{Bluhm:2004ep}
\begin{equation} \label{Eq:Bumblebee_action}
 S_{\text{bumb}}=S_B+S_{\partial \mathcal M}^{(\textrm{B})}+S_{\text{matter}}  \,,
\end{equation}
where
 \begin{subequations}
\begin{align} \label{Bumb_part1}
    S_B&=\int_{\mathcal{M}}\mathrm{d}^4x\,\sqrt{-g} \bigg[ \frac {1}{ 2\kappa} \big(  {R}+ \xi   
    B^\mu B^\nu       {R}_{\mu\nu} \big)-\frac{1}{4}B^{\mu\nu}B_{\mu\nu} \notag  
    \\ &\phantom{{}={}}\hspace{2.1cm}-V( B_{\mu}B^{\mu} - b^2 )\bigg]  \,,
\\
S_{\partial \mathcal M}^{(\textrm{B})}&=\oint_{\partial\mathcal{M}}   \mathrm{d} ^3 y  \, \varepsilon
  \frac{\sqrt q}{2\kappa}  \left[2K+\xi\Big(B^aB^b K_{ab} -\big(B^\mathbf{n}\big)^2 K\Big)   \right]  \,.
\end{align}
 \end{subequations}
\begin{figure}[b]
\centering
\includegraphics[scale=0.4]{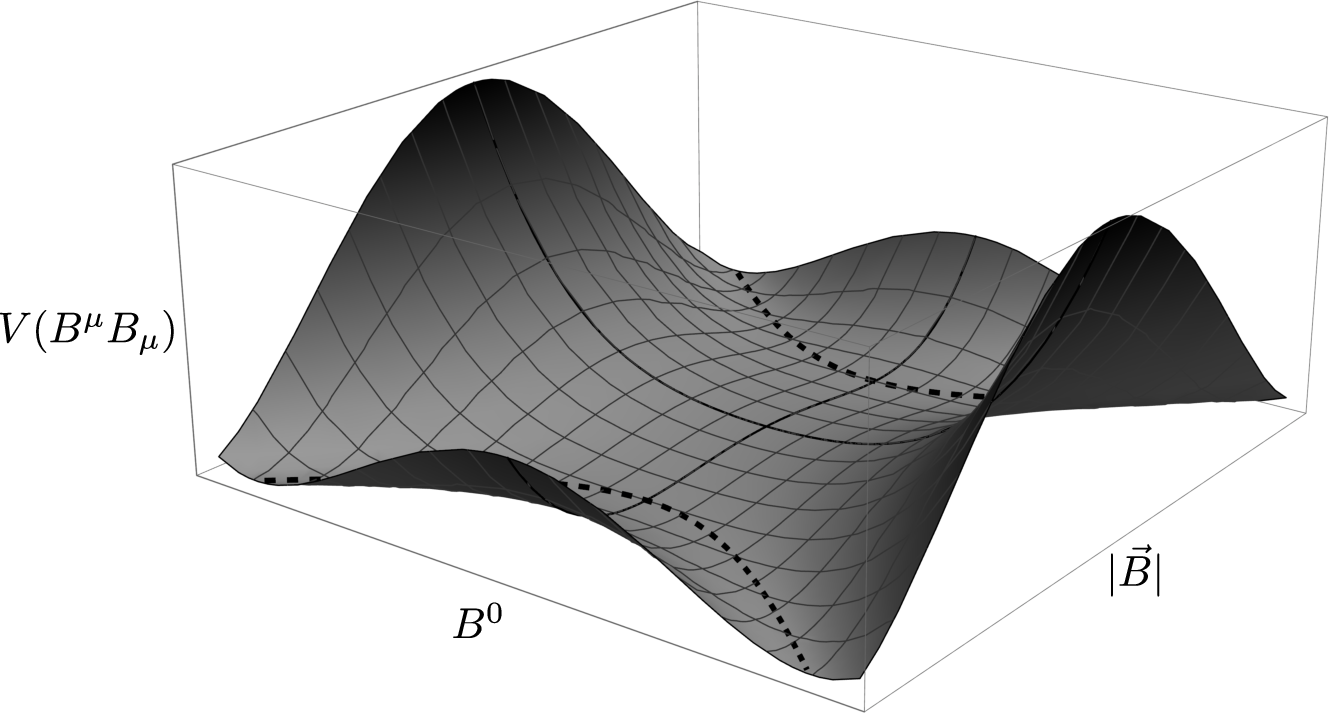}
\caption{Plot of the potential of the bumblebee field, which exhibits 
minima depicted by hyperbolas along the dashed lines. This structure arises 
from the spacetime metric signature and contrasts
the valley of minima in the Higgs potential, which is invariant under compact 
$\mathit{SU}(2)_L\otimes\mathit{U}(1)_Y$ transformations of the Higgs doublet.}
\label{Fig1}
\end{figure}
We define the field strength tensor of the bumblebee field as
$B_{\mu\nu}=\partial_\mu B_\nu-\partial_\nu B_\mu$ and assume that the bumblebee field acquires a vacuum expectation 
value, that is to say $\langle B_\mu \rangle =b_\mu$, in the potential
\begin{equation}\label{Potential}
    V(B_{\mu}B^{\mu} - b^2)=\frac{\lambda}{4} (B_{\mu}B^{\mu} - b^2 )  ^2\,,
\end{equation}
where $\lambda$ is a coupling constant; see Fig.~\refeq{Fig1}.
The modified Einstein equations read
\begin{equation}
\label{Einstein-bumblebee}
    G^{\mu\nu}=(T_B)^{\mu\nu}+ \kappa (T_m)^{\mu\nu}  \,,
\end{equation}
where 
\begin{align}\label{Energy-mom_bumblebee}
  (T_B)^{\mu\nu}&= \kappa\bigg[2V' B^\mu B^\nu+B^\mu_{\ \kappa}B^{\nu\kappa} -\bigg( V+\frac{1}{4}B^{\lambda\kappa}B_{\lambda\kappa} 
   \bigg)g^{\mu\nu}\bigg] \notag \\
   &\phantom{{}={}}+  \frac{\xi}{2} \bigg[B^\lambda B^\kappa R_{\lambda\kappa}g^{\mu\nu} -2\big(g^{\mu\rho}B^\nu +g^{\nu\rho}B^\mu\big) B^\sigma R_{\rho\sigma} \notag \\
   &\phantom{{}={}}\hspace{0.8cm}+\nabla_\lambda\nabla^\mu\big(B^\lambda B^\nu\big)+\nabla_\lambda\nabla^\nu \big(B^\lambda B^\mu\big) \notag \\
   &\phantom{{}={}}\hspace{0.8cm}-\nabla_\kappa\nabla_\lambda \big(B^\lambda B^\kappa
  \big)g^{\mu\nu} -\nabla_\lambda\nabla^\lambda\big( B^\mu B^\nu\big)   \bigg] \,,
\end{align}
and $(T_m)^{\mu\nu}$ is the energy-momentum tensor of matter.
The equation of motion of the bumblebee field is
\begin{equation} \label{bumblebee_field_eq}
    \nabla_\mu B^{\mu\nu}=2 V' B^\nu -\frac{\xi}{\kappa}B_\mu R^{\mu\nu}  \,,
\end{equation}
where the prime denotes differentiation with respect to the argument of $V$. 
\section{No-go constraint}
Let us consider the following gravitational action with explicit breaking:
\begin{equation}
\label{eq:modified-action}
S=\int\mathrm{d}^4x\,\frac{\sqrt{-g}}{2\kappa}\,\Big[R
+\mathcal{L}'(g_{\mu\nu},\bar{k}^{\alpha\beta\dots\omega})\Big]+S_{\partial \mathcal M}\,,
\end{equation}
where we have introduced the coefficients $\bar{k}^{\alpha\beta\dots\omega}$ of a generic nondynamical
background field.
By construction, the action is defined to be invariant under observer
diffeomorphisms, i.e., $\delta S_{\mathrm{obs}}=0$. However,
 it breaks particle diffeomorphisms such that $\delta S_{\mathrm{part}} \neq 0$; see Ref.~\refcite{Bluhm:2014oua}.
 
An observer transformation provides
\begin{align}\label{Obs_trans}
\delta S_{\mathrm{obs}}&=\int\mathrm{d}^4x\,\frac{\sqrt{-g}}{2\kappa}  \left[   \left(-G^{\mu\nu}+ 
T'^{\mu\nu}\right)\delta g_{\mu\nu}   +
J_{\alpha\beta\dots\omega}\delta \bar{k}^{\alpha\beta\dots\omega}     \right]   \,.
\end{align}
Contrarily, a particle transformation leads to
\begin{align}\label{Part_trans}
\delta S_{\mathrm{part}}&=\int\mathrm{d}^4x\,\frac{\sqrt{-g}}{2\kappa}  \left[   \left(-G^{\mu\nu}+ 
T'^{\mu\nu}\right)\delta g_{\mu\nu}     \right]   \,,
\end{align}
where we have defined
\begin{equation}
T'^{\mu\nu}=: \frac{1}{\sqrt{-g}} \frac{\delta(\sqrt{-g}\mathcal{L}')}{\delta g_{\mu\nu}}\,,\quad 
J_{\alpha\beta\dots\omega}=:\frac{\delta\mathcal{L}'}{\delta\bar{k}^{\alpha\beta\dots\omega}}\,.
\end{equation}
Performing several
integrations by parts in Eq.~\eqref{Obs_trans} and employing the 
geometric identities $\nabla^{\mu}G_{\mu\nu}=0$, we arrive at an identity for the background field coefficients:
\begin{align}
\label{eq:principal-equation-differential-form}
2\nabla_{\mu}T'^{\mu}_{\phantom{\mu}\,\,\nu}&=J_{\alpha\beta\dots\omega}
\nabla_{\nu}\bar{k}^{\alpha\beta\dots\omega}+\nabla_{\lambda}(J_{\nu\beta\dots\omega}\bar{k}^{\lambda\beta\dots\omega}) \notag \\
&\phantom{{}={}}+\nabla_{\lambda}(J_{\alpha\nu\dots\omega}\bar{k}^{\alpha\lambda\dots\omega})+\dots+\nabla_{\lambda}(J_{\alpha\beta\dots\nu}\bar{k}^{\alpha\beta\dots\lambda})\,.
\end{align}
For the theory being dynamically consistent, a critical requirement is that 
the following differential equation be satisfied:\cite{Kostelecky:2003fs,Kostelecky:2020hbb}
\begin{equation}
\label{eq:consistency-requirement}
\nabla_{\mu}T'^{\mu}_{\phantom{\mu}\,\,\nu}=0\,.
\end{equation}
Equations~\eqref{Obs_trans} and \eqref{Part_trans} imply the intriguing integral equation
\begin{equation}
\label{eq:principal-equation}
\delta S_{\mathrm{part}}+\int\mathrm{d}^4x\,\frac{\sqrt{-g}}{2\kappa}
J_{\alpha\beta\dots\omega}\delta \bar{k}^{\alpha\beta\dots\omega} =\delta S_{\text{obs}}  \,.
\end{equation}
In particular, an isometry generated by the Killing vector field $\chi$ satisfies $\delta\bar{k}^{\alpha\beta\dots\omega}
=\mathcal{L}_{\chi}\bar{k}^{\alpha\beta\dots\omega}=0$, where $\mathcal{L}_{\chi}$
is the Lie derivative along $\chi$. Then, Eq.~\eqref{eq:principal-equation} implies
$\delta S_{\mathrm{obs}}=0=  \delta S_{\mathrm{part}} $,
which restores\cite{Reyes:2024ywe} particle diffeomorphisms in the direction of $\chi$. 

In the next sections, we show that for a gravitational system that exhibits
Killing vectors fields it is possible to fulfill Eq.~\eqref{eq:consistency-requirement}.
Alternatively, one may restrict the space of metric solutions of the modified Einstein equations by imposing extra conditions
on spacetime geometry. For example, gravitational Chern-Simons 
theory\cite{Jackiw:2003pm} can be treated accordingly by requiring that ${}^{*}RR=0$ with the dual Riemann tensor ${}^{*}R$.
\section{Cosmology: Isotropy and homogeneity}
We start from a cosmological setting with the Friedmann metric 
\begin{equation}
   \mathrm{d}s^2=-\mathrm{d}t^2 + a^2(t) \left[ \frac{\mathrm{d}r^2}{1-kr^2}
    +r^2  \left( \mathrm{d}\theta^2+\sin^2\theta \mathrm{d}\phi^2     \right)  \right ]  \,,
\end{equation}
and consider the presence of a background which can be chosen to violate spacetime symmetries either explicitly or spontaneously.
The modified Einstein equation are 
 \begin{equation}\label{E-eq}
    G_{\mu\nu}+\Lambda g_{\mu\nu}= {T}_{\mu\nu}+\kappa(T_m)_{\mu\nu}\,,
\end{equation}
where $T_{\mu \nu}$ depends on the background fields and $(T_m)_{\mu\nu}$ describes energy-momentum of a perfect fluid.

Let us consider the transformations along the Killing directions $\chi$
characterizing isotropy and homogeneity and apply them to Eq.~\eqref{E-eq}. Then,
\begin{equation}\label{form-invariant-em}
    \mathcal{L}_{\chi} {T}_{\mu\nu}=0 \,. 
\end{equation}
From the examples given in Ref.~\refcite{Reyes:2024hqi}, we have 
found that in the specific sectors $u,s$ and $t$, the above condition leads to  
\begin{equation} \label{Form-Invariant}
\mathcal{L}_{\chi }u=\mathcal{L}_{\chi } s^{\mu\nu}
=\mathcal{L}_{\chi} t^{\mu\nu\rho\sigma}=0\,.
\end{equation}
For spontaneous breaking in the bumblebee model, the Lie derivative of the bumblebee field is not restricted,
but only its spatial norm. This procedure simplifies the differential 
equations of Eq.~\eqref{eq:consistency-requirement}, which allows us to find nontrivial solutions of the backgrounds compatible with the dynamics.
\section{Friedmann equations and cosmic acceleration}
This section is dedicated to applying the method of Killing vectors to both dynamical and nondynamical 
background fields. We focus on the $t$ sector. Results on the $u$ and $s$ sectors including the Hubble tension
can be consulted in Refs.\  \refcite{Reyes:2022dil,Khodadi:2023ezj}.
\subsection{The t background}
The first step to analyze cosmology in the presence of backgrounds is to 
perform the ADM decomposition.\cite{Arnowitt:1962hi,ONeal-Ault:2020ebv,Reyes:2021cpx}
The decomposition of spacetime $\mathcal M(x^{\mu})$ 
relies on the foliation in terms of a family of hypersurfaces
$\Sigma_t(y^a)$ of constant time $t$.
According to this method of decomposing spacetime, we define the projector 
$e^\mu_a:=\frac{\partial x^{\mu}}{d y^a}  $.
The analysis simplifies when considering a purely tangential background $t$. Hence, we introduce the projections
$ t^{\mu\nu\rho\sigma}=e^\mu_ae^\nu_be^\rho_ce^\sigma_d t^{abcd}$.
After applying the method to decompose the modified Einstein equations,
from Eq.~\eqref{EM} without cosmological constant,
the first modified Friedmann equation can be written as
\begin{align}\label{FE1}
H^2 &=\frac{1}{3\Big(1-\frac{1}{3}q_{ab} q_{cd}  t^{acbd}\Big)}  \bigg( \kappa \rho -\frac{k}{a(t)^2}
q_{ab}q_{cd}t^{acbd}    -\frac{3k}{a(t)^2}\bigg)  \,,
\end{align}
and the second modified Friedmann equation reads
\begin{align}\label{FE2}
\dot{H}+H^2&=\frac{-1}{6\Big(1-\frac{1}{3}q_{ab}q_{cd}t^{acbd}\Big)}\bigg[\kappa (\rho+3P) 
+\frac{4k}{a(t)^2}q_{ab}q_{cd} t^{acbd} \notag \\
&\phantom{{}={}}\hspace{1cm}-2q_{ac}\Big( D_b D_d t^{abcd}+Hq_{bd}\dot{t}^{abcd}   +2H^2q_{bd}t^{abcd}\Big) \bigg]  \,,
 \end{align}
where $H(t)=\dot a(t) /a(t)$ is the Hubble parameter and $D_a$ the induced covariant derivative on the hypersurface, and $k$ the spatial curvature. 
The matter source is described by a perfect fluid of energy density $\rho$
and pressure $P$ with equation of state $P= w \rho$.

We follow the method previously explained to obtain 
a form-invariant background field; see Eq.~\eqref{Form-Invariant}. The result for 
spatially flat $k=0$ is
\begin{eqnarray}
    t^{abcd}=a(t)^4\eta  \big(q^{ac}q^{bd}-q^{ad}q^{bc}\big) \,,
\end{eqnarray} 
with a constant $\eta$.
\begin{figure}[b]
	\centering
	\includegraphics[scale=0.71]{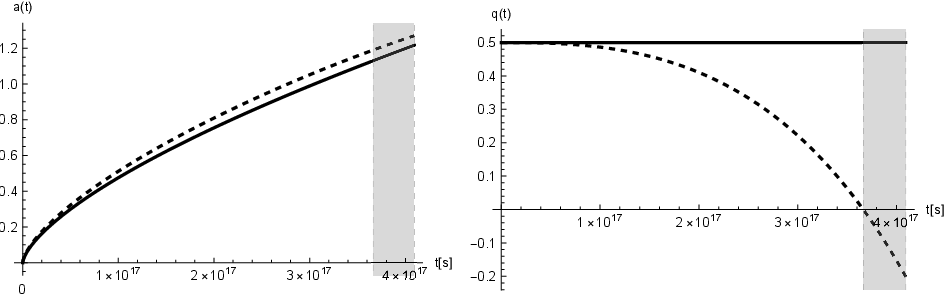}
	\caption{Scale factor $a(t)$ and deceleration parameter $q(t)$ for the choice of $w=0$.
 The solid lines show the standard behaviors and the dashed lines the 
 solutions of the modified Friedmann equations~\eqref{FE1} and \eqref{FE2}. The 
 gray areas illustrate the time intervals when accelerated expansion occurs.}
	\label{Fig3}
\end{figure}
The deceleration parameter $q=-\ddot{a}a/\dot{a}^2$ is employed for the analysis.
Configurations ought to be found that satisfy the condition $q<0$. The latter is the case when $1 - 2 \eta a^4 > 0$ 
and $(1 + 3 w) - 2 (5 + 3 w) \eta a^4 < 0$, which can be summarized as
\begin{equation}
\label{conditionsacc}
\left( \frac{1 + 3 w}{5 + 3 w} \right) \frac{1}{2 \eta} < a^4 < \frac{1}{2 \eta}\,.
\end{equation}
We solve Eqs.~\eqref{FE1} and \eqref{FE2} numerically by considering only matter ($w=0$) to get an idea about the behavior 
of the scale factor. In this case, the relationship $\rho_0=3H_0\Omega_m/\kappa$ for the fluid density is valuable. It holds 
that $\Omega_m=1$, since we analyze matter only. Moreover, the present value 
$H_0$ for the Hubble parameter determined by the Planck Collaboration\cite{Planck:2018vyg} is chosen. As an example,
$\eta=5\times 10^{-2}$ is taken, which leads to the plots for $a$ and $q$ displayed in the left- and right-hand-side panels, respectively, 
of Fig.~\ref{Fig3}. For the purpose of comparison, the standard behavior is presented, too.
\subsection{The bumblebee field}
For the bumblebee field with action~\eqref{Bumb_part1}, we consider a purely tangential 
time-dependent bumblebee field 
\begin{align}
   B^\alpha=e^\alpha_{a}B^a (t) \,.
\end{align}
The modified Einstein equations~\eqref{Einstein-bumblebee} imply the first and second modified Friedmann equations. In particular,
\begin{subequations}
\begin{align}
H^2 &= \frac{1}{3\Big(1-\frac{\xi}{3}B^cB_c\Big)}\bigg[\kappa \rho+\kappa\bigg( V+\frac{1}{2}  q^{cd}\dot{B}_c\dot{B}_d 
 \bigg) \notag \\
 &\phantom{{}={}}\hspace{2.6cm}-\frac{\xi}{2}\bigg(\frac{2k}{a(t)^2}B^c B_c  +2HB^c\dot{B}_c\bigg)-\frac{3k}{a(t)^2}\bigg]\,,
\end{align}
and
\begin{align}
  \dot{H}+H^2&= \frac{-1}{6\Big(  1  -\frac{\xi}{3}B^cB_c\Big)}\bigg[\kappa (3P+\rho)+\kappa\Big(   2V'B^cB_c-2V+q^{cd}\dot{B}_c\dot{B}_d  \Big) \nonumber \\
       &\phantom{{}={}}\hspace{1cm}+\frac{\xi}{2}\bigg(-\frac{4k}{a(t)^2}B^c B_c+4H^2B^c B_c -10Hq^{cd}\dot{B}_cB_d \notag \\
       &\phantom{{}={}}\hspace{1.9cm}+2q^{cd}\ddot{B}_cB_d+2q^{cd}\dot{B}_c\dot{B}_d\bigg)\bigg] \,.
\end{align}
\end{subequations}
The bumblebee field equations are
\begin{align}
  \ddot{B}_a+ H\dot{B}_a+2 \bigg[V' -\frac{\xi}{2\kappa}\bigg(\frac{2k}{a(t)^2} + \dot{H}+3H^2\bigg)\bigg]B_a=0  \,.
\end{align}
The central point of the above equations is that we have found possible configurations of the bumblebee field that respect the symmetries of
isotropy and homogeneity by demanding $ (T_B)^{\mu\nu}$ of the bumblebee field to fulfill 
Eq.~\eqref{form-invariant-em}.
Notably,
the arising configuration for the bumblebee field restricts the value of $B^aB_a$, but not the direction of $B^a$.
Studying possible applications in this context is left for the future.
\section{Conclusions}
Here, we have compiled some of the critical findings of a series of recent 
works.\cite{Reyes:2021cpx,Reyes:2024hqi,Reyes:2024ywe} Scenarios of explicit 
diffeomorphism violation that avoid the no-go constraints have been presented. The 
modified Einstein equations were decomposed along the directions of the ADM 
decomposition to arrive at the modified Friedmann equations. Interestingly, regimes of 
accelerated cosmological expansion can arise without a cosmological constant. For the 
ADM decomposition of the bumblebee model and a more detailed discussion of 
cosmological phenomenology, the reader can consult Ref.~\refcite{Reyes:2024hqi}.
\section*{Acknowledgments}
CMR acknowledges support from Fondecyt~1241369. CR acknowledges support from
 the ANID fellowship No.~21211384 and Universidad de Concepci\'on. 
  MS is indebted to CNPq Produtividade~310076/2021-8 and CAPES/Finance Code 001.

\end{document}